\documentclass[journal]{IEEEtran}

\usepackage{caption}
\usepackage{clrscode}
\usepackage{bbm}
\usepackage{amsfonts}
\usepackage[dvips]{graphicx}
\usepackage{times}
\usepackage{cite}
\usepackage{amsmath}
\usepackage{array}
\usepackage{amssymb}

\usepackage{stfloats}
\usepackage{graphicx}
\usepackage{subfigure}
\usepackage{footnote}
\usepackage{amsthm}
\usepackage{booktabs}
\usepackage{array}
\usepackage{algorithm}
\usepackage{algorithmic}
\usepackage{subeqnarray}
\usepackage{cases}
\usepackage{threeparttable}
\usepackage{color}
\usepackage{multirow}
\usepackage{xcolor}
\usepackage{epstopdf}
\usepackage{bm}
\usepackage{dsfont}
\usepackage{enumerate}
\usepackage{float}
\usepackage{hyperref}

\begin{document}
\title{On the Fundamental Limits of MIMO Massive \\ Multiple Access Channels}

\author{\IEEEauthorblockN{Fan Wei, Yongpeng Wu, Wen Chen, Wei Yang, and Giuseppe Caire \\}

\thanks{The work of Y. Wu is supported in part by  the National Science Foundation (NSFC) under Grant 61701301
and Young Elite Scientist Sponsorship Program by CAST.
The work of W. Chen is supported in part by NSFC under Grant $61671294$, STCSM Project under Grant $16$JC$1402900$ and $17510740700$, National Science and Technology Major Project under Grant $2018$ZX$03001009-002$. The corresponding author of this paper is Y. Wu.}

\thanks{F. Wei and W. Chen are with Shanghai Institute of Advanced Communications and Data Sciences, the Department of Electronic
Engineering, Shanghai Jiao Tong University, Minhang 200240, China (E-mail: weifan89@sjtu.edu.cn; wenchen@sjtu.edu.cn).}

\thanks{Y. Wu is with the Department of Electronic Engineering, Shanghai Jiao Tong University,
Minhang 200240, China (e-mail: yongpeng.wu@sjtu.edu.cn).}

\thanks{W. Yang is with Qualcomm Research, Qualcomm Technologies, Inc., San Diego, CA, $92121$, USA (e-mail: weiyang@qti.qualcomm.com)}

\thanks{G. Caire is with Institute for Telecommunication Systems, Technical University Berlin, Einsteinufer 25,
10587 Berlin, Germany (Email: caire@tu-berlin.de). }

}
\maketitle

\begin{abstract}
In this paper, we study the multiple-antenna wireless communication networks, where a large number of devices simultaneously communicate with an access point. The capacity region of multiple-input multiple-output massive multiple access channels (MIMO mMAC) is investigated. While joint typicality decoding is utilized to establish the achievability of capacity region for conventional MAC with fixed number of users, the technique is not directly applicable for MIMO mMAC. Instead, an information-theoretic approach based on Gallager's error exponent analysis is exploited to characterize the \textcolor[rgb]{0,0,0}{finite dimension region} of MIMO mMAC. Theoretical results reveal that the finite dimension region of MIMO mMAC is dominated by sum rate constraint only, and the individual user rate is determined by a specific factor that corresponds to the allocation of sum rate. The rate in conventional MAC is not achievable with massive multiple access, which is due to the fact that successive interference cancellation cannot guarantee an arbitrary small error decoding probability for MIMO mMAC. The results further imply that, asymptotically, the individual user rate is independent of the number of transmit antennas, and channel hardening makes the individual user rate close to that when only statistic knowledge of channel is available at receiver. The finite dimension region of MIMO mMAC is a generalization of the symmetric rate in Chen \emph{et al.} (2017).
\end{abstract}

\IEEEpeerreviewmaketitle

\section{Introduction}
Massive connectivity and massive capacity are two central requirements for future cellular network. In contrast to the current human-centric networks, the expansion of connected machine-centric networks will potentially support millions of devices in each cellular network. As such, it is vital to understand the fundamental limits of such communication systems. The capacity region of conventional multiple access channel (MAC) where the number of simultaneous communicate devices is fixed has been established~\cite{01a,01b,01}. However, the proof of achievability by joint typicality may not be suitable when the number of devices in the network grows unbounded with code block length~\cite{02}. \textcolor[rgb]{0,0,0}{Moreover, with fixed transmit powers as well as fixed number of antennas, the individual user rate would approach zero in the massive connectivity network, which makes the traditional notion of rate defined by $R=\log M / n\rightarrow0$, i.e., bits per channel use a less meaningful performance metric.}

In order to circumvent the above issues, Chen \emph{et al.}~\cite{02,03} proposed a symmetric code construction for the massive multiple access channels (mMAC), where the number of devices grows unbounded with the code block length. \textcolor[rgb]{0,0,0}{Meanwhile, the message-length capacity, which is a codelength related notion of capacity was introduced. By the definition of message-length capacity, the individual user rate now is different from conventional vanishing rate defined by bits per channel use, and is ensured to be bounded away from zero.}

In this paper, the region of multiple-input multiple-output massive multiple access channels (MIMO mMAC) is investigated. We adopt the definition of message-length capacity from~\cite{02,03} in order to avoid zero individual user rate. \textcolor[rgb]{0,0,0}{Asymmetric code construction, where different user groups may have codes with different rates is considered in this paper. Therefore the devices can transmit at diverse rates depending on their Quality of Service. The number of message set size is assumed to be finite and denumerable. Thereby, a finite dimension region of MIMO mMAC is formulated in this paper.} Moreover, to meet the massive capacity requirement in the future cellular network, multiple-antenna system~\cite{04,05,Wu2016TIT,Wu2018P} is considered in our work.

In the massive connectivity regime, when the number of users grows to infinity with the code block length, the joint typicality method fails to achieve the region of MIMO mMAC due to the exponential number of error events. As such, we investigate the region of MIMO mMAC based on Gallager's error exponent analysis. In contrast to the conventional capacity region that is a polytope defined in general by $2^{K} -1$ rate constraint inequalities, where $K$ is the number of users, the region of MIMO mMAC is shown to be subjected to the sum rate constraint only and each individual use rate can be expressed in terms of a specific factor multiplied by the sum rate. Theoretical analysis of the individual rate shows that asymptotically, there is no benefit of having multiple antenna at the transmitters and what matters is the number of antennas at the receiver. Further, we show that due to the channel hardening effect, the asymptotic individual user rate converges to that when receiver has the statistical knowledge of channels only. The result has suggested that the instant estimation for users' channel gain at receiver (which is unrealistic in the context of massive accessed users) is unnecessary. Another interesting difference between the conventional MAC and the case studied in this paper is that, while successive interference cancellation (SIC) is able to achieve rate points on the capacity region boundary in conventional case, it fails in MIMO mMAC due to the arbitrary large interference from the other users.

For the related works of massive multiple access, the upper bound of sum rate in multiple antenna system is provided in~\cite{06} but with a sketch proof only. The key difference between Yu's work in~\cite{06} and mMAC is that there the code block length goes to infinity before the  number of devices goes to infinity. A compressed sensing based method is proposed for the active users detection in MIMO random users access~\cite{07}. It was shown that the probability of miss detections and false alarms vanishes with an infinity number of antennas. In~\cite{08}, the achievability bound for the random-access code is proposed and the tradeoff between the energy-per-bit and spectral efficiency for var,ious MAC regimes in the massive multiple access is studied. Moreover, some recent works on the coding aspects of massive random access can be found in~\cite{08b,08c,08d,08f,Wei2018VTM,Wei2019TWC}.

\emph{Notation}: Lowercase letters $x$, bold lowercase letters $\mathbf{x}$, and bold uppercase letters $\mathbf{X}$ denote scalars, column vectors, and matrices, respectively. We use $(\cdot)^{\dagger}$ to denote complex conjugate and transpose for matrix and $\mbox{Tr}\{\cdot\}$ to denote the trace operator of matrix. The notion $\binom{n}{k}$ denotes the binomial coefficient of $n$ choose $k$. The binary entropy function is denoted by $H_{2}(p)=-p\log p-(1-p)\log(1-p)$.


\section{System Model}
We consider the MIMO mMAC system where each user is equipped with $N_{T}$ antennas while $N_{R}$ antennas are deployed at the base station. In the massive multiple access, the number of users is comparable or even larger than the code block length, therefore we denote by $k_{n}$ the number of users, which grows linearly with the code block length $n$, i.e., $k_{n}=O(n)$~\cite{03,08}. For the $i_{th}$ channel use, the received signal can be written as
\begin{equation}\label{01}
  \mathbf{y}(i) = \sum_{k=1}^{k_{n}}\mathbf{H}_{k}(i)\mathbf{x}_{k}(i)+\mathbf{z}(i), \texttt{ } i=1,2,\ldots,n,
\end{equation}
where $\mathbf{H}_{k}(i) \in \mathcal{C}^{N_{R}\times N_{T}}$ and $\mathbf{x}_{k}(i) \in \mathcal{C}^{N_{T}\times 1}$ are the complex channel matrix and transmitted codewords vector of user $k$, respectively. $\mathbf{z}(i) \in \mathcal{C}^{N_{R}\times 1}$ is the complex white Gaussian noise vector with normalized identity covariance matrix. Without loss of generality, we assume for every user $k$ and every channel use $i$, the entries of channel matrix $\mathbf{H}_{k}(i)$ equal to a fast fading coefficient multiplied by a large scale fading factor~\cite{05}, i.e.,
\begin{equation}\label{01a}
  h_{m,n}^{(k)} = \alpha_{m,n}^{(k)}\beta_{k}^{\frac{1}{2}},
\end{equation}
where $\alpha_{m,n}^{(k)}$ denotes the fast fading coefficient, which is assumed to be complex Gaussian with zero mean and unit variance. The coefficient $\beta_{k}=z_{k}/ r_{k}^{\eta}$ is large scale fading factor where $r_{k}$ is the distance between user $k$ and the base station, $\eta$ is the path loss exponent, and $z_{k}$ is the log-normal random variable with a standard deviation of $\sigma_{\mathrm{shad}}$. Both $r_{k}$ and $z_{k}$ are statistically independent of the antenna indexes but are statistically dependent on user index $k$. We also assume that the large scale fading factors remain constant during the transmission of codewords. Meanwhile, the Gaussian noise $\mathbf{z}(i)$ are supposed to be independent and identically distributed (\emph{i.i.d.}) in each channel use $i$.

With the transmissions over $n$ channel uses, the uplink received signal is given by
\begin{equation}\label{02}
  \mathbf{y}^{n} = \sum_{k=1}^{k_{n}}\mathbf{H}^{n}_{k}\mathbf{x}^{n}_{k}+\mathbf{z}^{n},
\end{equation}
where $\mathbf{y}^{n} = \big[\mathbf{y}^{T}(1), \mathbf{y}^{T}(2),\cdots ,\mathbf{y}^{T}(n)\big]^{T}$ is the $nN_{R} \times 1 $ received signal. The block diagonal channel matrix and the transmitted codewords vector for each user $k$ is given by $\mathbf{H}^{n}_{k} = \mathrm{diag}\big\{\mathbf{H}_{k}(1),\mathbf{H}_{k}(2),\cdots, \mathbf{H}_{k}(n)\big\}$ and $\mathbf{x}^{n}_{k} = \big[\mathbf{x}^{T}_{k}(1),\mathbf{x}^{T}_{k}(2), \cdots, \mathbf{x}^{T}_{k}(n)\big]^{T}$, respectively. The $nN_{R} \times 1$ vector $\mathbf{z}^{n} = \big[\mathbf{z}^{T}(1), \mathbf{z}^{T}(2),\cdots ,\mathbf{z}^{T}(n)\big]^{T}$ denotes the Gaussian noise. In the following of this paper, the variables are associated with superscript $n$ when collections of channel uses are considered, whereas the superscript is removed to denote the per channel use variables.

The following definitions introduce the asymmetric codes construction for MIMO mMAC as well as the notion of message-length capacity~\cite{03}.

\emph{Definition 1 :}  Let $\mathcal{X}_{k}$ and $\mathcal{Y}$ be the alphabet of input symbol for user $k$ and output symbol of channel, respectively. An $(M_{1}, M_{2},\ldots, M_{k_{n}},n)$ code for MIMO mMAC $(\mathcal{X}_{1} \times \mathcal{X}_{2} \times ... \times \mathcal{X}_{k_{n}}, P_{Y|X_{1},...,X_{k_{n}}}, \mathcal{Y})$ consists of $k_{n}$ sets of integers $\mathcal{W}_{k} \in \{1,2,...,M_{k}\}$ called \emph{message sets} and

\begin{itemize}
  \item The set of \emph{encoding functions} $\mathcal{E}_{k}: \mathcal{W}_{k}\rightarrow \mathcal{X}^{nN_{T}}_{k} $  for every user $k$. Since all users are active, the transmitted codewords $\mathbf{x}_{k}$ should satisfy the following power constraint:
\begin{equation*}
  P_{k}-\delta \leq \mbox{Tr}\big(\mathbf{Q}_{k}\big) \leq P_{k},
\end{equation*}
where $\delta>0$ is an arbitrary small number and $\beta_{k}P_{k} \in [P_{\min}, P_{\max}]$. The positive semi-definite matrix $\mathbf{Q}_{k} = \mathbb{E}\big\{\mathbf{x}_{k}\mathbf{x}_{k}^{\dag}\big\} $ denotes the input covariance matrix for user $k$. For user $k$, the symbol $\mathbf{x}_{k}$ are assumed to be generated \emph{i.i.d.} according to the Gaussian distribution $\mathcal{CN}(\mathbf{0};\mathbf{Q}_{k})$.
  \item A \emph{decoding function} $\mathcal{D}: \mathcal{Y}^{nN_{R}} \rightarrow \mathcal{W}_{1} \times \mathcal{W}_{2} \times ... \times \mathcal{W}_{k_{n}} $, which is a deterministic rule that assigns a sequence of estimated messages $(\mathcal{W}_{1}, \mathcal{W}_{2}, ... ,\mathcal{W}_{k_{n}})$ to each possible received vector.
\end{itemize}

The average error probability of the $(M_{1},M_{2},\ldots, M_{k_{n}},n)$ code is given by
\begin{equation}\label{03}
  P_{e}^{(n)} = \mathrm{Pr}\big\{D(\mathbf{y}^{n}) \neq (\mathcal{W}_{1}, \mathcal{W}_{2}, \ldots ,\mathcal{W}_{k_{n}} )\big\},
\end{equation}
where the messages $(\mathcal{W}_{1}, \mathcal{W}_{2}, \ldots ,\mathcal{W}_{k_{n}})$ are generated independently over the message set.

\emph{Definition 2 (Asymptotically Achievable Message-Length):} Given the set of functions $R_{k}(\cdot)$ that map natural numbers to some positive value, we say the message-length $R_{k}(n)$ are asymptotically achievable if there exists a sequence of $(\lceil\exp(R_{1}(n))\rceil,\lceil\exp(R_{2}(n))\rceil,\ldots, \lceil\exp(R_{k_{n}}(n))\rceil,n)$ codes in the sense of Definition $1$ such that the average error probability $P_{e}^{(n)}$ vanishes as $n \rightarrow \infty$.

As stated in~\cite{01,03}, when the total number of senders is very large, the traditional notion of rate, measured in bits per channel use, goes to zero for each individual user even the total amount of information can be arbitrary large. Thus the message-length rate, as a function of the codelength $n$, is adopted here to avoid the deficiencies of conventional capacity.

\textcolor[rgb]{0,0,0}{The conventional definition for the capacity region of MIMO mMAC is the closure of the set of all achievable rate tuples $(R_{1}(n),R_{2}(n),\ldots,R_{k_{n}}(n))$. However, as $n$ becomes larger, the dimension of this rate tuple also increases since the number of users now grows linearly with $n$. Thus, to avoid an increasing dimension of capacity region and by noting that the message set size $M_{k}$ may not necessarily different from each other, we define the finite dimension region of MIMO mMAC as follows.}

\textcolor[rgb]{0,0,0}{\emph{Definition 3 (Finite Dimension Message-Length Region):} Let $K_{j}$ denote the number of users with message set size $\exp\{V_{j}\}$, i.e., with message-length rate $R_{k}=V_{j}$ for $k\in \mathcal{S}$, where $V_{j}$ are the set of ``distinct'' message-length rates in the system. Then by grouping together the rates $R_{k}$ that are equal to each other, the $J$-dimensional region of MIMO mMAC is defined as the closure of the convex hull of all $(K_{1},K_{2},\ldots, K_{J})$, such that $\sum_{j=1}^{J}K_{j}V_{j}$ is upper bounded by the sum rate of the system.}

\textcolor[rgb]{0,0,0}{For the finite and denumerable message set size, the dimension $J$ was expected to be independent of codelength $n$. Thus, in \emph{Definition 3}, a finite dimension region is formulated through the number of sustainable users for some given rate.}

\textcolor[rgb]{0,0,0}{For the presentation of the main results in next section, we further define the coefficients $\mu_{k}^{n} = \frac{\log M_{k}}{\sum_{t \in \mathcal{S}}\log M_{t}} \in (0,1)$, given the message set size $M_{k}$ of each user, and assume that the inverse of $\mu_{k}^{n}$ grows linearly with codelength $n$, i.e., $\frac{1}{\mu_{k}^{n}}=O(n)$ so that $\lim \limits _{n\rightarrow\infty} n\mu_{k}^{n}  = c_{k} $, where $c_{k}$ is some positive constant such that $\sum_{k=1}^{k_{n}}c_{k} = n$.}


\section{The Finite Region of MIMO mMAC}
In this section, we present the main results of this paper. We assume that the channel distribution information is available at the transmitter (CDIT) only. Besides, the users are supposed to be simultaneously active in the network \footnote{The random multiple access where each user transmits with a certain probability (as the case in~\cite{07}) will be treated in the future work.}.

\emph{Theorem 1:} For MIMO mMAC described in~\eqref{02}, let $\mathcal{S}= \{1,2, \ldots, k_{n}\}$ denote the user set in the network, where the total number of users scales as $k_{n}=O(n)$. Then message-length vector with components $(R_{1},\ldots, R_{k_n})$ is asymptotically achievable if
\begin{equation}\label{04}
R_{k} \leq c_{k}\mathbb{E}_{\mathbf{H}}\Big\{\log\det\Big(\mathbf{I}+\sum \limits_{t \in \mathcal{S}}\mathbf{H}_{t}\mathbf{Q}_{t}\mathbf{H}^{\dagger}_{t}\Big)\Big\},
\end{equation}
where $\mathbf{H} = (\mathbf{H}_{1},\mathbf{H}_{2},...,\mathbf{H}_{k_{n}})$ with $\mathbf{H}_{k}$ denotes the channel matrix of user $k$, and $c_{k}$ is defined as in Section II. \hfill $\square$

\emph{Corollary:} With massive accessed users, since $k_{n}=O(n)$, the asymptotic individual message-length rate grows as
\begin{align}
 \label{05a}R_{k} \xrightarrow{|\mathcal{S}|\rightarrow\infty} &\,c_{k}N_{R}\log\Big(1+\sum_{t\in\mathcal{S}}\beta_{t}P_{t}\Big)\\
 \label{05b}=&\,c_{k}N_{R}O(\log n)+O(1),
\end{align}
where the term $O(\log n)$ is independent of any other constants but relates only to $n$, and the $O(1)$ constant depends on the pathlosses and transmit powers. \hfill $\square$

What is important here is the $\log(n)$ dependence for the individual rate. This means that the number of bits that each user is able to deliver grows logarithmically with the block length and therefore also logarithmically with the number of users.

\textcolor[rgb]{0,0,0}{Notice that for each user $k$, rate~\eqref{04} is determined by $c_{k}$ (or $M_{k}$). Since the message set size $M_{k}$ may not necessarily different from each other, given the \emph{Definition 3}, the finite dimension region of MIMO mMAC is formulated as
\begin{align}\label{05}
  \nonumber &\mathcal{C}_{\mathcal{MAC}}  \triangleq \Bigg\{(K_{1},...,K_{J}):
  K_{1}V_{1}+K_{2}V_{2}
  +\cdots\\&+K_{J} V_{J} \leq n\mathbb{E}_{\mathbf{H}}\Big\{\log\det\Big(\mathbf{I}+\sum \limits_{t \in \mathcal{S}}\mathbf{H}_{t}\mathbf{Q}_{t}\mathbf{H}^{\dagger}_{t}\Big)\Big\}\Bigg\},
\end{align}
where $J$ denotes dimension of the region, and $V_{j}$ is on the order of $O(\log n)$.} Eq.~\eqref{05} states that any numbers of sustainable users are feasible, provided the total operational rate is constrained by the sum capacity of the system.

$\mathbf{Remark \, 1}: $ The individual user rate in~\eqref{04} is different from that of conventional MAC. In conventional MAC, the corner points of the region is achieved through SIC, i.e., the receiver decodes the first user's data by treating the signals from others as noises and then subtract it from the received signal. The SIC approach, however, may not be possible in the massive connectivity scenario since interferences plus noise become arbitrarily large as the number of users $k_{n}$ grows without bound. \footnote{A mathematical analysis for the single antenna AWGN case can be found in~\cite{03} and the multiple antennas case will be analyzed in the future work.} As shown in \emph{Theorem 1}, the region of MIMO mMAC is only subjected to the sum rate constraint and the individual user rate is obtained by the sum rate multiplies a factor $c_{k}$, which is related to the size of message set $M_{k}$ as shown in Section II. Note that the size $M_{k}$ is determined when the codebook has been generated. This is in contrast with conventional MAC where the region is subjected to $2^{k_{n}}-1$ constraints. $\hfill \blacksquare$

$\mathbf{Remark \, 2}: $ From~\eqref{05a}, the multiple antennas have provided the system with the degree of freedom gain~\cite{12}, which is $N_{\mathrm{DoF}} = \min\{N_{R},k_{n}N_{T}\}=N_{R}$. One important implication of~\eqref{05a} is that, no matter how many transmitter antennas each user is equipped with, we will get the same asymptotical result. Thus there is no benefit of having multiple antenna at the transmitters and what matters is the number of antennas at the receiver. The result in~\eqref{05a} also indicates that due to the effect of channel hardening, the asymptotic rate is close to that when the receiver has only the statistic knowledge of effective channel gain. As a consequence, the estimation of users' fast fading coefficients $\alpha_{m,n}^{(k)}$ at receiver (which may be unrealistic due to the massive number of users) is unnecessary. $\hfill \blacksquare$

\section{Proof of the Converse}
The proof of the converse part is mainly based on the Fano's inequality. We have to show that for any sequence of codes $(M_{1},M_{2},\ldots, M_{k_{n}},n)$ with $P^{(n)}_{e} \rightarrow 0$ must have $R_{k}=\log M_{k} \leq c_{k}\mathbb{E}_{\mathbf{H}}\big\{\log\det\big(\mathbf{I}+\sum_{t \in \mathcal{S}}\mathbf{H}_{t}\mathbf{Q}_{t}\mathbf{H}^{\dagger}_{t}\big)\big\} $ for some positive constant $c_{k}$.

Let $\mathbf{W}=\big\{\mathcal{W}_{1},\mathcal{W}_{2},\ldots, \mathcal{W}_{k_{n}}\big\}$ denote the message of $k_{n}$ users. The entropy of messages are computed by
\begin{align}\label{06}
  \nonumber H(\mathbf{W})& = H(\mathbf{W}|\mathbf{y}^{n})+I(\mathbf{W};\mathbf{y}^{n})  \\
            & \leq H(\mathbf{W}|\mathbf{y}^{n})+I(\mathbf{x}^{n};\mathbf{y}^{n}),
\end{align}
where $\mathbf{X}^{n}= \big\{\mathbf{X}^{n}_{1},\mathbf{X}^{n}_{2},\ldots, \mathbf{X}^{n}_{k_{n}}\big\}$ and~\eqref{06} follows from the data processing inequality since $\mathbf{W}\rightarrow \mathbf{x}^{n} \rightarrow \mathbf{y}^{n}$ forms a Markov chain.

The mutual information in~\eqref{06} can be written as,
\begin{align}\label{07}
  \nonumber I(\mathbf{x}^{n};\mathbf{y}^{n}) & = H(\mathbf{y}^{n})-H(\mathbf{y}^{n}|\mathbf{x}^{n})  \\
  \nonumber   & = H(\mathbf{y}^{n})-H(\mathbf{z}^{n}) \\
  \nonumber   & \leq \sum_{i=1}^{n}\Big[H(\mathbf{y}(i))- H(\mathbf{z}(i))\Big] \\
              & \leq n\mathbb{E}_{\mathbf{H}} \Big\{\log\det\Big(\mathbf{I}+\sum \limits_{t \in \mathcal{S}}\mathbf{H}_{t}\mathbf{Q}_{t}\mathbf{H}_{t}^{\dagger}\Big)\Big\},
\end{align}
where the second inequality follows because the entropy is maximized by normal distribution with the covariance $\mathbf{K}_{\mathbf{y}} = \mathbf{I}+\sum_{t \in \mathcal{S}}\mathbf{H}_{t}\mathbf{Q}_{t}\mathbf{H}_{t}^{\dagger}$.

By Fano's inequality, the conditional entropy $H(\mathbf{W}|\mathbf{y}^{n})$ is upper bounded by
\begin{equation}\label{08}
  H(\mathbf{W}|\mathbf{y}^{n}) \leq 1+P^{(n)}_{e}\sum\limits_{k \in \mathcal{S}}\log M_{k}.
\end{equation}

For the uniformly distributed messages, the entropy is given by $H(\mathbf{W}) = \sum_{k \in \mathcal{S}}\log M_{k} $. Combining~\eqref{06},~\eqref{07}, and~\eqref{08}, we obtain
\begin{align}\label{09}
  \nonumber \sum\limits_{k \in \mathcal{S}}\log M_{k} \leq  n\mathbb{E}_{\mathbf{H}}
            \Big\{&\log\det\Big(\mathbf{I}+\sum \limits_{t \in
            \mathcal{S}}\mathbf{H}_{t}\mathbf{Q}_{t}\mathbf{H}_{t}^{\dagger}\Big)\Big\} \\
            & + 1 + P^{(n)}_{e}  \sum\limits_{k \in \mathcal{S}} \log  M_{k}.
\end{align}
Multiplying both sides with $\mu_{k}^{n}$,
\begin{align}\label{10}
  \nonumber (1-P^{(n)}_{e})\log M_{k} &\leq \mu_{k}^{n} \\
  + n\mu_{k}^{n} \mathbb{E}_{\mathbf{H}}& \Big\{\log\det\Big(\mathbf{I}+\sum \limits_{t \in
            \mathcal{S}}\mathbf{H}_{t}\mathbf{Q}_{t}\mathbf{H}_{t}^{\dagger}\Big)\Big\}.
\end{align}
Now let $n \rightarrow \infty$, the first term on R.H.S. of~\eqref{10} vanishes. As a consequence, the rate $R_{k}$ is shown to be
\begin{equation}\label{11}
 \log M_{k} \leq c_{k}\mathbb{E}_{\mathbf{H}}\Big\{\log\det\Big(\mathbf{I}+\sum \limits_{t \in
            \mathcal{S}}\mathbf{H}_{t}\mathbf{Q}_{t}\mathbf{H}_{t}^{\dagger}\Big)\Big\},
\end{equation}
for some positive constant $c_{k}$.

\section{Proof of the Achievability}
In this section, we give the proof of achievability part for MIMO message-length capacity. The joint typicality decoding is shown to be achievable in conventional multiple access channel. This approach, however, fails in the case of MIMO mMAC as $k_{n}$ now grows linearly with the codelength $n$. As a consequence, the packing lemma~\cite{08a} does not hold in this regime. An alternative approach for the achievability part is by Gallager's error exponent analysis~\cite{09,10}, which forms the basis of our proof in this section. In the following, we assume that the decoder uses \emph{maximum likelihood} detection for the data decoding.

Consider the case when $k_{e}$ out of $k_{n}$ users incur a decoding error. Clearly, there are $\binom{k_{n}}{k_{e}}$ such error events. Based on the Gallager's error exponent analysis, the decoding error probability for the $l_{th}$ event is given by~\cite{09}
\begin{align}\label{12}
  \nonumber P^{l}_{e}(k_{e}) & \leq \prod \limits_{k \in A_{l}}(M_{k}-1)^{\rho}\mathbb{E}_{\mathbf{H}}\Bigg\{
   \int Q\{\mathbf{X}_{A_{l}^{c}}\} \\
   \nonumber \times\Bigg[&\int Q\{\mathbf{X}_{A_{l}}\}
   P^{\frac{1}{1+\rho}}\{\mathbf{y}|\mathbf{X},\mathbf{H}\}d\mathbf{X}_{A_{l}} \Bigg]^{1+\rho} d\mathbf{X}_{A_{l}^{c}}d\mathbf{y}\Bigg\}^{n} \\
  & \leq \exp\Bigg\{-\Bigg[nE^{l}(\rho,Q)-\rho \sum\limits_{k \in A_{l}}R_{k}\Bigg]\Bigg\},
\end{align}
where $\rho \in [0,1]$, $\mathbf{X}=\{\mathbf{x}_{1},\mathbf{x}_{2},\ldots,\mathbf{x}_{k_{n}}\}$, and $A_{l}\subseteq \mathcal{S}$ is the $l_{th}$ subset of $\mathcal{S}$ with $|A_{l}|=k_{e}$ while $A_{l}^{c}$ is the complementary set of $A_{l}$. The error exponent $E^{l}(\rho,Q)$ is given by
\begin{align}\label{13}
  \nonumber E^{l}(\rho,Q) & = -\log\mathbb{E}_{\mathbf{H}}\Bigg\{\int Q\{\mathbf{X}_{A_{l}^{c}}\} \\
  \times \Bigg[\int & Q\{\mathbf{X}_{A_{l}}\}
  P^{\frac{1}{1+\rho}}\{\mathbf{y}|\mathbf{X},\mathbf{H}\}d\mathbf{X}_{A_{l}} \Bigg]^{1+\rho} d\mathbf{X}_{A_{l}^{c}}d\mathbf{y}\Bigg\},
\end{align}
where $Q\{\cdot\}$ denotes a $priori$ distribution of transmitted codewords and $P\{\mathbf{y}|\mathbf{X},\mathbf{H}\}$ is the likelihood function for each channel use. Assume the a $priori$ distribution of the transmitted codewords $Q\{\cdot\}$ follows a Gaussian distribution $\mathcal{CN}(\mathbf{0};\mathbf{Q}_{k})$, after some algebra, Eq.~\eqref{13} can be calculated as (analogous to the results in~\cite[p. 592]{04})
\begin{equation}\label{14}
  E^{l}(\rho) = -\log \mathbb{E}_{\mathbf{H}} \Big\{
  \det\Big(\mathbf{I}+\frac{1}{1+\rho}\sum \limits_{t \in A_{l}}
  \mathbf{H}_{t}\mathbf{Q}_{t}\mathbf{H}^{\dagger}_{t}\Big)^{-\rho}\Big\}.
\end{equation}

By union bound, the $k_{e}$-users decoding error probability is upper bounded by
\begin{align}\label{15}
   P_{e}(k_{e}) & \leq \sum_{A_{l} \subseteq \mathcal{S}} P_{e}^{l}(k_{e}).
\end{align}
Since there are $\binom{k_{n}}{k_{e}}$ error decoding events, using the inequality~\cite[Section 12.1]{01}
\begin{equation}\label{15a}
  \binom{k_{n}}{k_{e}} \leq \exp\big(k_{n}H_{2}(\gamma)\big),
\end{equation}
where $\gamma = k_{e}/k_{n}$, we have for arbitrary $l$,
\begin{align}\label{16}
   \binom{k_{n}}{k_{e}}P_{e}^{l}(k_{e}) \leq \exp\Big\{-\Big[E_{o}(\rho)-k_{n}H_{2}(\gamma)\Big]\Big\},
\end{align}
where $E_{o}(\rho)=nE^{l}(\rho,Q)-\rho \sum_{k \in A_{l}}R_{k}$.

Let $E_{r}(\rho) = E^{l}(\rho)-\frac{\rho}{n}\sum_{k \in A_{l}}R_{k} -\frac{k_{n}}{n}H_{2}(\gamma)$ denote the per channel user error exponent, we have the following lemma.

\emph{Lemma 1:} For $\epsilon > 0$, if the message-length rate $R_{k}$ is given by
\begin{equation}\label{17}
  R_{k} = (1-\epsilon)c_{k}\mathbb{E}_{\mathbf{H}}\Big\{\log\det\Big(\mathbf{I}+\sum \limits_{t \in \mathcal{S}}\mathbf{H}_{t}\mathbf{Q}_{t}\mathbf{H}^{\dagger}_{t}\Big)\Big\},
\end{equation}
for the constants $c_{k}>0$ such that $\sum_{k \in \mathcal{S}}c_{k}=n$, and assume the maximum transmit power $P_{max}$ is upper bounded by some constant. Then there exists a positive constant $c_{0}>0$ such that,
\begin{equation}\label{18}
  E_{r}(\rho) \geq c_{0},
\end{equation}
holds for all sufficient large codelength $n$.

\emph{Proof:}
By Gallager's $\rho$ trick~\cite{10}, $\rho$ can be an arbitrary real number within the interval $[0,1]$ in the error exponent analysis. Without loss of generality, we choose $\rho=1$ in the proof. Combining~\eqref{14} and~\eqref{17}, the error exponent is given by
\begin{align}\label{19}
  \nonumber E_{r}(1,\gamma) = & - \log \mathbb{E}_{\mathbf{H}} \Big\{\det\Big(\mathbf{I}+\frac{1}{2}\sum \limits_{t \in A_{l}}\mathbf{H}_{t}\mathbf{Q}_{t}\mathbf{H}^{\dagger}_{t}\Big)^{-1}\Big\} \\
  \nonumber -\frac{(1-\epsilon)}{n}&\sum_{k \in A_{l}}c_{k}\mathbb{E}_{\mathbf{H}} \Big\{\log \det\Big(\mathbf{I}+\sum \limits_{t \in \mathcal{S}}\mathbf{H}_{t}\mathbf{Q}_{t}\mathbf{H}^{\dagger}_{t}\Big)\Big\}\\
   &-\frac{k_{n}}{n}H_{2}(\gamma).
\end{align}
In the following, we consider two cases and show that $E_{r}(1,\gamma)>c_{0}$ holds for sufficient large $n$.

\emph{Case I:} $k_{e}$ is a large number and grows linearly with the code length $n$ so that $\lim \limits_{n\rightarrow \infty} \frac{k_{e}}{k_{n}} = \gamma > 0$ converges to some constant. Since the summation $\frac{1}{n}\sum_{k \in A_{l}}c_{k}$ is less than $1$ and $H_{2}(\gamma) < 1$, the error exponent is lower bounded by
\begin{align}\label{20}
  \nonumber E_{r}(1,\gamma) \geq & \,\epsilon \mathbb{E}_{\mathbf{H}}\Big\{\log\det\Big(\mathbf{I}+\sum \limits_{t \in\mathcal{S}}\mathbf{H}_{t}\mathbf{Q}_{t}\mathbf{H}^{\dagger}_{t}\Big)\Big\}-\frac{k_{n}}{n}\\
  \nonumber -\Bigg\{\log &\mathbb{E}_{\mathbf{H}} \Big[
   \det\Big(\mathbf{I}+\frac{1}{2}\sum \limits_{t \in A_{l}}\mathbf{H}_{t}\mathbf{Q}_{t}\mathbf{H}^{\dagger}_{t}\Big)^{-1}\Big] \\
   +&\mathbb{E}_{\mathbf{H}}\Big[\log\det\Big(\mathbf{I}+\sum \limits_{t \in \mathcal{S}}\mathbf{H}_{t}\mathbf{Q}_{t}\mathbf{H}^{\dagger}_{t}\Big)\Big]\Bigg\}.
\end{align}

We now define $\mathbf{G}_{t} = \mathbf{H}_{t}\mathbf{Q}_{t}\mathbf{H}^{\dagger}_{t}$ where the $(i,j)_{th}$ entry is given by
\begin{align}\label{26}
  \nonumber g_{i,j}^{(t)} &= \sum_{n}\sum_{m}h_{i,m}^{(k)}q_{m,n}^{(t)}(h_{j,n}^{(t)})^{*} \\
   &=\sum_{n}\sum_{m}\alpha_{i,m}^{(k)}q_{m,n}^{(t)}(\alpha_{j,n}^{(t)})^{*}\beta_{t}
\end{align}
where $h_{m,n}^{(t)}=\alpha_{m,n}^{(t)}\beta_{t}^{\frac{1}{2}}$ and $q_{m,n}^{(t)}$ denote the $(m,n)_{th}$ entry of matrices $\mathbf{H}_{t}$  and $\mathbf{Q}_{t}$ , respectively. Since the fast fading coefficients $\alpha_{m,n}^{(t)}$ are \emph{i.i.d.} random variables with normal distribution $\mathcal{CN}(0,1)$ , by Kolmogorov's strong law of large number~\cite{11},
\begin{equation}\label{27}
  \lim \limits_{k_{n} \rightarrow \infty} \sum\limits_{t \in \mathcal{S}}g_{i,j}^{(t)} = \left\{
                             \begin{array}{ll}
                               \sum\limits_{t \in \mathcal{S}}\beta_{t}\mathrm{Tr}(\mathbf{Q}_{t}), & \hbox{$i = j$;} \\
                               0, & \hbox{$i \neq j$.}
                             \end{array}
                           \right.
\end{equation}
and
\begin{equation}\label{28}
  \lim \limits_{k_{e} \rightarrow \infty} \sum\limits_{t \in A_{l}}g_{i,j}^{(t)} = \left\{
                             \begin{array}{ll}
                               \sum\limits_{t \in A_{l}}\beta_{t}\mathrm{Tr}(\mathbf{Q}_{t}), & \hbox{$i = j$;} \\
                               0, & \hbox{$i \neq j$.}
                             \end{array}
                           \right.
\end{equation}
Therefore for lager number $k_{e}$ and $k_{n}$,
\begin{equation}\label{29}
  \lim \limits_{k_{n}\rightarrow \infty} \det\Big(\mathbf{I}+\sum \limits_{t \in \mathcal{S}}\mathbf{G}_{t}\Big)
= \Big(1+  \sum\limits_{t \in \mathcal{S}}\beta_{t}\mathrm{Tr}(\mathbf{Q}_{t})\Big)^{N_{R}},
\end{equation}
and
\begin{equation}\label{30}
  \lim \limits_{k_{e} \rightarrow \infty} \det\Big(\mathbf{I}+\frac{1}{2}\sum \limits_{t \in A_{l}}\mathbf{G}_{t}\Big)
= \Big(1+\frac{1}{2}\sum\limits_{t \in A_{l}}\beta_{t}\mathrm{Tr}(\mathbf{Q}_{t})\Big)^{N_{R}}.
\end{equation}
Consequently,
\begin{align}\label{31}
  \nonumber E_{r}(1,\gamma) &\geq  \epsilon {N_{R}} \log \Big(1+ \sum\limits_{t \in \mathcal{S}}\beta_{t}\mathrm{Tr}(\mathbf{Q}_{t})\Big) -\frac{k_{n}}{n} \\
  \nonumber +N_{R} \log &\Big\{
  \Big(1+ \frac{1}{2}\sum\limits_{t \in A_{l}}\beta_{t}\mathrm{Tr}(\mathbf{Q}_{t})\Big) \Big / \Big(1+ \sum\limits_{t \in \mathcal{S}}\beta_{t}\mathrm{Tr}(\mathbf{Q}_{t})\Big)\Big\}\\
  \nonumber &\geq \epsilon {N_{R}} \log \Big(1+ \sum\limits_{t \in \mathcal{S}}\beta_{t}\mathrm{Tr}(\mathbf{Q}_{t})\Big) -\frac{k_{n}}{n}\\
  +N_{R}\log &\Big\{\Big(1+ \frac{1}{2}\gamma k_{n} (P_{\min}-\delta)\Big)\Big /\Big(1+ k_{n}P_{\max}\Big)\Big\},
\end{align}
where the last inequality in~\eqref{31} follows because $P_{\min}-\delta=\min_{t\in A_{l}}\beta_{t}\mathrm{Tr}(\mathbf{Q}_{t})$ and $P_{\max}=\max_{t\in \mathcal{S}}\beta_{t}\mathrm{Tr}(\mathbf{Q}_{t})$. Since the last two terms in~\eqref{31} converge to some constant while the first term grows unbounded as $n$
increases, the error exponent is shown to be large than some positive constant.

\emph{Case II:} $k_{e}$ scales sublinearly with codelength $n$ so that $\lim \limits_{n\rightarrow \infty} \gamma
= \lim \limits_{n\rightarrow \infty} \frac{k_{e}}{k_{n}} = 0$. By choosing $\rho=1$, we have
\begin{align}\label{32}
  \nonumber E_{r}(1,\gamma) = & - \log \mathbb{E}_{\mathbf{H}} \Big\{\det\Big(\mathbf{I}+\frac{1}{2}\sum \limits_{t \in A_{l}}\mathbf{H}_{t}\mathbf{Q}_{t}\mathbf{H}^{\dagger}_{t}\Big)^{-1}\Big\} \\
  \nonumber -\frac{(1-\epsilon)}{n}&\sum_{k \in A_{l}}c_{k}\mathbb{E}_{\mathbf{H}} \Big\{\log \det\Big(\mathbf{I}+\sum \limits_{t \in \mathcal{S}}\mathbf{H}_{t}\mathbf{Q}_{t}\mathbf{H}^{\dagger}_{t}\Big)\Big\}\\
   &-\frac{k_{n}}{n}H_{2}(\gamma).
\end{align}

\textcolor[rgb]{0,0,0}{Note that since matrix $\mathbf{H}_{t}\mathbf{Q}_{t}\mathbf{H}^{\dagger}_{t}$ is positive semi-definite, its largest eigenvalue is strictly positive, and hence the largest eigenvalue of $\mathbf{I}+\frac{1}{2}\sum_{t \in A_{l}}\mathbf{H}_{t}\mathbf{Q}_{t}\mathbf{H}^{\dagger}_{t}$ is strictly larger than $1$, whereas the rest eigenvalues can be greater than or equal to $1$. As a consequence, the first term in~\eqref{32} is strictly positive.}

For the second term, we have $\frac{1}{n}\sum_{k \in A_{l}}c_{k} \leq k_{e}c_{\max}/ n$. Since $c_{k}$ is a positive constant for all $k$ while the expectation in the second term is on the order $O(\log n)$. In the following, we consider two cases for the sublinearly growing $k_{e}$:
\begin{enumerate}
  \item $k_{e} = O(\frac{n}{\log n})$, i.e., $k_{e}$ is on the order of $\frac{n}{\log n}$,
  \item $k_{e} = o(\frac{n}{\log n})$, i.e., $k_{e}$ grows slower than the speed $\frac{n}{\log n}$.
\end{enumerate}
In case $1$), the first term grows unbounded with the codelength, the second term is lower bounded by some constant and the last term vanishes. In case $2$), the last two terms vanishes while the first term is positive. Therefore, in both cases the error exponent is shown to be large than zero.

Since $l$ is arbitrary, by~\eqref{15}, \emph{Lemma 1} and combining the \emph{Case I} and \emph{Case II}, the error probability $P_{e}(k_{e})$ decays exponentially with codelength $n$. Hence,
\begin{equation}\label{33}
  P_{e} \leq \sum_{k=1}^{k_{n}} P_{e}(k_{e})\leq k_{n}e^{-nc_{0}},
\end{equation}
vanishes as $n$ increases. As a consequence, the achievability of \emph{Theorem 1} is established. The corollary can be obtained by using the same metric in the proof of \emph{Case I}.

\section{Numerical results}
This section presents some numerical results for the MIMO mMAC channels. We consider a single cell communication system with $k_{n}$ users randomly deployed within the cellular network. Each user has a transmit power $P_{k}$ generated uniformly within the interval $[5,15]$. The distance between an individual user and the base station ranges from $100$m to $1000$m. For large scale fading, the decay exponent is $\eta=3.8$ whereas the shadow-fading standard deviation is set to be $\sigma_{\mathrm{shad}}=8.0$ dB.

\begin{figure}[t]
  \centering
  \includegraphics[width=3.0in, height=2.5in]{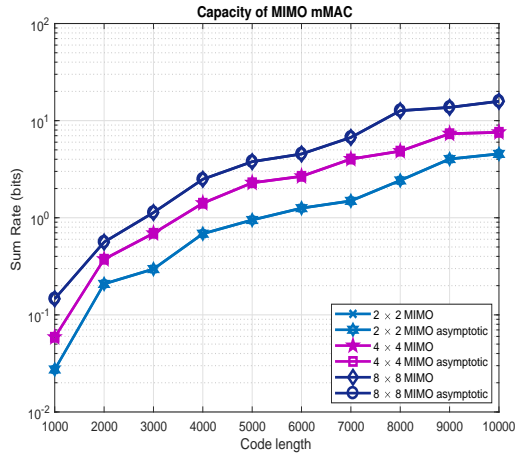}
  \caption{The sum rate of MIMO mMAC versus growing codelength $n$.}\label{Fig.2}
\end{figure}

In Fig.~\ref{Fig.2}, we plot the sum rate of MIMO mMAC system as a function of codelength $n$. The considered MIMO systems include $2 \times 2$, $4 \times 4$ and $8 \times 8$ MIMO, respectively. To meet the massive connectivity requirement, the number of users is set $k_{n}=n/2$, which grows linearly with the codelength $n$. Note that since $k_{n}$ grows unbounded with codelength $n$ and the message-length capacity considered in this paper itself is a function of $n$, the capacity of MIMO mMAC system thereby grows with an increasing of codelength $n$ as shown in Fig.~\ref{Fig.2}. Moreover, it can be also observed from Fig.~\ref{Fig.2} that the asymptotic rate~\eqref{05a} matches well with the operational rate~\eqref{04}. Hence, with \emph{i.i.d.} Rayleigh coefficients, the system performance is only affected by large scale fading under the massive connectivity scenario. The results in Fig.~\ref{Fig.2} indicate that with massive number of users, the estimation of users' instant channel gain is unnessary due to the channel hardening.

\begin{figure}[t]
  \centering
  \includegraphics[width=3.0in, height=2.5in]{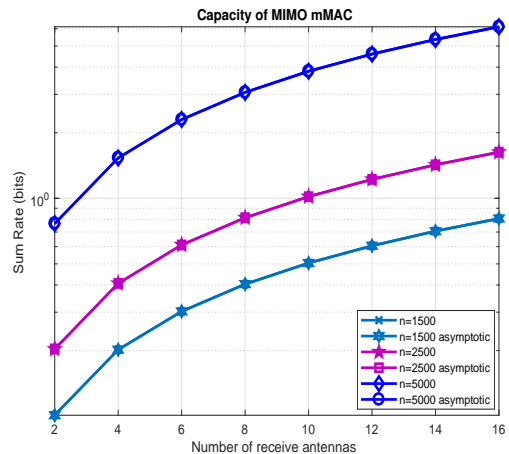}
  \caption{The sum rate of MIMO mMAC versus growing receive antennas $N_{R}$, $N_{T}=2$.}\label{Fig.3}
\end{figure}

We also demonstrate the impact of receive antenna numbers on the capacity of MIMO mMAC system in Fig.~\ref{Fig.3}. We observe from Fig.~\ref{Fig.3} that given different codelength $n$, the capacities all grow linearly with an increasing of $N_{R}$. This is due to the degree of freedom gain of MIMO system~\cite{12}. The similar phenomenon has been well observed in~\cite{13}, where they show that the number of degree of freedom of a MIMO system is limited by the minimum of the number of transmit antenna and the number of receive antenna in high SNR regime. With massive users, the uplink SNR is high enough. Therefore, we have $N_{\mathrm{DoF}} = \min\{N_{R},k_{n}N_{T}\}=N_{R}$. As a result, the capacity of MIMO mMAC grows linearly with $N_{R}$ as is shown in Fig.~\ref{Fig.3}.

\section{Conclusion}
In this work, we have investigated the capacity of mMAC with multiple antennas. The capacity of conventional MAC with fixed number of users is achieved by a joint typicality decoder at the receiver. However, as the number of users grows unbounded, the joint typicality approach does not work due to the exponential number of error events. As such, we derived the finite dimension region of MIMO mMAC channels by using the Gallager's error exponent analysis. Theoretical analysis indicates that rather than the interference plus noise, the individual user rate in MIMO mMAC is determined by a specific factor that corresponds to the allocation of sum capacity.

In this paper, we assumed that the set of active users is known to the receiver. In a random access scenario, each user has an independent transmit probability and the receiver is lack of the active users knowledge. This generalized model will be addressed in the future work.

\end{document}